\documentclass[a4paper,twocolumn,amsmath,amssymb,floatfix,prl,preprintnumbers,footinbib]{revtex4}
\usepackage{graphicx}
\usepackage{dcolumn}
\usepackage{bm}

\begin{document}


\title{Observation of Excited States in a Graphene Quantum Dot}
\author{S.~Schnez, F.~Molitor, C.~Stampfer, J.~G{\"u}ttinger, I.~Shorubalko, T.~Ihn, and K.~Ensslin}      

\affiliation{Solid State Physics Laboratory, ETH Z\"urich, 8093 Z\"urich, Switzerland}

\begin{abstract}
We demonstrate that excited states in single-layer graphene quantum dots can be detected via direct transport experiments. Coulomb diamond measurements show distinct features of sequential tunneling through an excited state. Moreover, the onset of inelastic co-tunneling in the diamond region could be detected. For low magnetic fields, the position of the single-particle energy levels fluctuate on the scale of a flux quantum penetrating the dot area. For higher magnetic fields, the transition to the formation of Landau levels is observed. Estimates based on the linear energy-momentum relation of graphene give carrier numbers of the order of 10 for our device.
\end{abstract}

\maketitle


Graphene \cite{Geim07, Neto07}, the first real two-dimensional (2D) solid, consists of a hexagonal lattice of carbon atoms providing highly mobile electrons \cite{Novoselov05, Zhang05} for future applications in electronics, spintronics \cite{Tombros07} and information processing \cite{Trauzettel07}. However, confinement of charge carriers in graphene cannot be achieved as easily as in conventional two-dimensional electron gases by using electrostatic gates because of the gapless nature of graphene \cite{Neto07} and a relativistic phenomenon called Klein tunneling \cite{Dombey99, Katsnelson06}. Cutting graphene into a desired geometry is an alternative to overcome this obstacle. Well-controlled nanostructures, such as nanoribbons \cite{Chen07, Han07, Li08}, quantum interference devices \cite{Miao08, Russo08, Molitor08}, and single-electron transistors \cite{Stampfer08, Ponomarenko08, Stampfer08b} have been created in several labs to date. Small spin-orbit and hyperfine interactions have been theoretically predicted \cite{Min06}, promising spin decoherence times superior to the GaAs material system in which solid-state spin qubits are most advanced today \cite{Petta05, Koppens06}. Therefore, the identification of individual orbital quantum states, well established in GaAs quantum dot devices, have so far remained on the wish list of physicists aiming at quantum information processing with graphene. 

An atomic force microscope image of our quantum dot (QD) is shown in Fig. 1a. It was fabricated with the standard procedure: Mechanical exfoliation of natural graphite led to single-layer graphene flakes. The desired structure was defined with electron beam lithography and subsequently cut using reactive ion etching based on Ar and O$_2$. Contacts were also defined with electron beam lithography; then gold contacts were evaporated on top \cite{Stampfer08}. The single layer quality was experimentally verified with Raman spectroscopy \cite{Graf07}. The QD device consists of two about 60 nm and 70 nm wide graphene constrictions separating source (S) and drain (D) contacts from the graphene island (diameter 140 nm). The island can be tuned by a nearby plunger gate (PG), whereas the overall Fermi level is adjusted with a highly doped silicon back gate (BG). The sample was annealed for about 24 hours at $400\,\textrm{K}$ directly before cool down. The experiments were carried out in a dilution refrigerator at a base temperature of 40 mK. Measuring the current $I$ through the QD as a function of back gate voltage $V_{\textrm{bg}}$ allows us to identify a transport gap \cite{Stampfer08} extending roughly from $V_{\textrm{bg}} = -8\,\textrm{V}$ to $8\,\textrm{V}$ (Fig. 1b). Since the gap is centered around zero back gate voltage, we have little doping of our graphene device. Characteristic peaks in the gap-region were identified as Coulomb peaks (Fig. 1c) and could be used to estimate an upper bound for the electronic temperature. This was found to be around 200 mK. In the following measurements, we set the back gate voltage to zero in order to tune the device close to the charge neutrality point.

\begin{figure}
  \centering
  \includegraphics[width=.45\textwidth]{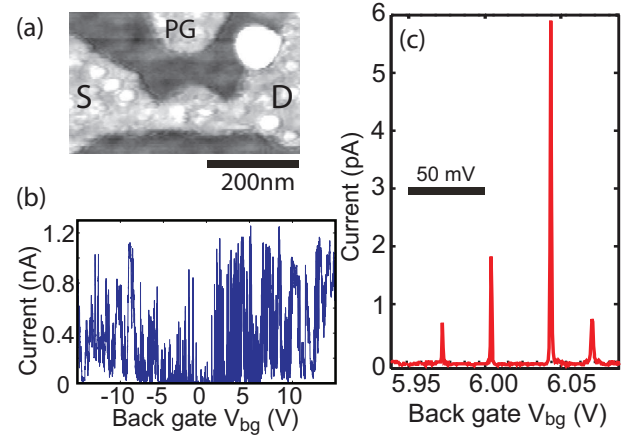}
  \caption{(Color online) (a) Atomic force micrographs of the measured quantum dot. The quantum dot can be tuned by a nearby plunger gate (PG). The central island is connected to source (S) and drain (D) contacts by two constrictions. The diameter of the dot is 140 nm. (b) A backgate sweep shows a transport gap from roughly $V_{\textrm{bg}} = -8\,\textrm{V}$ to $8\,\textrm{V}$ (bias voltage of $V_{\textrm{bias}}=3\,\textrm{mV}$). (c) Coulomb blockade measured with the back gate at an electronic temperature of 200 mK and a bias voltage of $V_{\textrm{bias}} = 16\,\mu\textrm{V}$.}
\end{figure}

Coulomb diamond measurements \cite{Kouwenhoven97}, i.e., plots of the differential conductance $G = \textrm{d}I/\textrm{d}V_{\textrm{bias}}$, as a function of the quantum dot bias voltage $V_{\textrm{bias}}$ and plunger gate voltage $V_{\textrm{pg}}$ are shown in Fig. 2. Within this plunger gate voltage range, no charge rearrangements were observed and the sample was stable for more than two weeks. We extract a typical energy scale of the order of 10 meV. A strong fluctuation of the addition energy over the plunger gate voltage range $-0.1\,\textrm{V} < V_{\textrm{pg}} < 1.2\,\textrm{V}$ (full data range not shown), corresponding to an energy range of around 100 meV is observed, indicating the importance of quantum confinement effects.

\begin{figure}
  \centering
  \includegraphics[width=.45\textwidth]{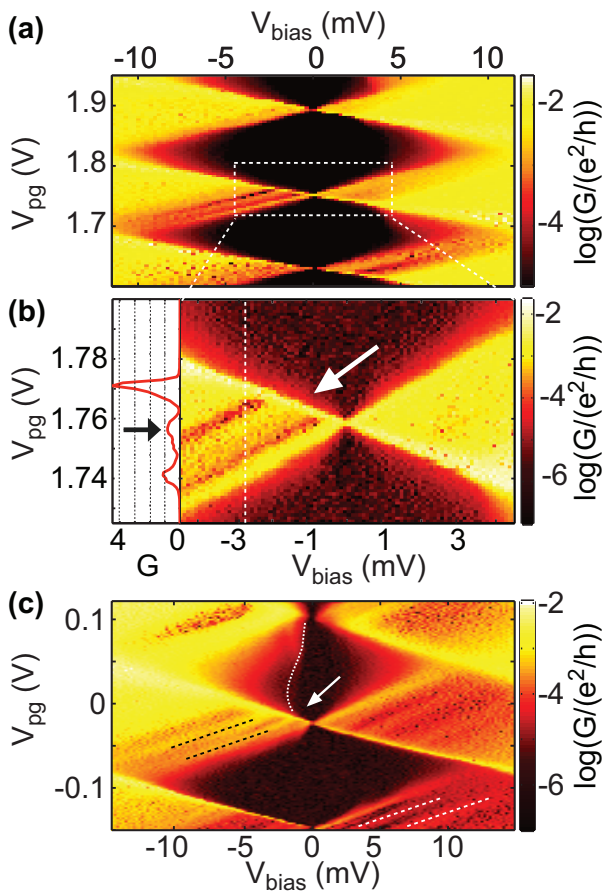}
  \caption{(Color online) (a) Differential conductance $G$ (logarithmic plot) as a function of source-drain voltage $V_{\textrm{bias}}$ and plunger gate voltage $V_{\textrm{pg}}$. (b) The right panel is a zoom of the enframed area in (a). An excited state is clearly visible (white arrow). A cut along the dashed line at $V_{\textrm{bias}} = -2.87\,\textrm{mV}$ is shown in the left panel (here $G$ is measured in units of $10^{-3} e^2/h$ and was smoothed over 4 points). (c) Stability diagram at different plunger gate voltages. Several excited states are visible as shown by dashed lines. In the upper diamond, regions of higher conductance (in the left part of the diamond) can be seen. This is interpreted as a signature of co-tunneling in a graphene quantum dot (see arrow). In all measurements shown in Figs. 2 and 3, the back gate voltage was set to $V_{\textrm{bg}} = 0\,\textrm{V}$ and the electronic temperature was around 200 mK as deduced from the Coulomb peak width.}
\end{figure}

This is supported by the observation of excited states, which appear in Fig. 2a as distinct lines of increased conductance running parallel to the edge of the Coulomb diamonds \cite{Kouwenhoven97}. Fig. 2b – showing a close up of Fig. 2a – allows to extract an excitation energy $\Delta \approx 1.6\,\textrm{meV}$ as marked by the white arrow. A line cut at $V_{\textrm{bias}} = − 2.78\,\textrm{mV}$ (dashed line) presented in the left panel of Fig. 2b shows the peak of the excited state at finite bias (arrow). The broadening of the peak significantly exceeds thermal broadening and might be due to the energy dependent coupling of the excited state to the graphene leads.

Fig. 2c shows two Coulomb diamonds at lower plunger gate voltage, where more than one excited state is observed as a function of increasing energy, as shown by pairs of dashed lines. These excitations are found at energies of around 1.6 meV and 3.3 meV (black dashed lines) and 2.1 meV and 4.2 meV (white dashed lines), respectively. The observation of excitations at finite source-drain voltage finds support by the detection of inelastic co-tunneling onsets at lower bias. Inside the upper Coulomb diamond of Fig. 2c, we distinguish between regions of suppressed and slightly elevated conductance separated by the dotted line. The edge of this conductance step is aligned with the (first) excited state outside the diamond at an energy of 1.6 meV as highlighted by an arrow.

The number of charge carriers on the quantum dot can be estimated by using the linear density of states of graphene $D(E) = 2E/(\pi v_F^2\hbar^2)$, where $v_F \approx 10^6\,\textrm{m/s}$ is the Fermi velocity \cite{Neto07}. We find for the single-particle level spacing $\Delta(N) = \hbar v_F/(d\sqrt{N})$ \footnote{This estimation does not hold for very small $N$, in particular not for $N = 0$. However, it is a reasonable estimation for electron numbers exceeding $N \geq 5$.}, where $N$ is the number of charge carriers and $d$ is an effective dot diameter related to the dot area $A$ according to $A=\pi d^2/4$. From the measured excitation energy and the lithographic dimensions of the dot ($d = 140\,\textrm{nm}$) we estimate the number of charge carriers on the dot to be of the order of 10. This estimation obviously changes if the dot area is larger, e.~g. if the dot was not defined by the two constrictions. The number of charge carriers would then be even lower. However, magnetic field sweeps corroborate our estimation of the dot size as pointed out below. The current was below measurement resolution at $V_{\textrm{pg}} < -0.5\,\textrm{V}$ so that smaller charge carrier numbers and the potential electron-hole crossover could not be studied. The Coulomb diamonds shown in Fig. 2c scatter significantly stronger in size than those presented in Fig. 2a. This might be a consequence of the lower number of charge carriers on the quantum dot. According to theoretical considerations \cite{Schnez08}, electron-hole symmetry could manifest itself by an enhanced confinement energy. This should be detectable by the type of experiment described here.

Using typical values for the addition energy and excitation energies we estimate the charging energy to be $\sim 8.5\,\textrm{meV}$, which agrees reasonably well with the energy estimated from a disk model $\Delta E_C = e^2/(4\epsilon \epsilon_{\textrm{eff}}d) \approx 12\,\textrm{meV}$. We assumed the effective dielectric constant including vacuum and the $\textrm{SiO}_2$ to be $\epsilon_{\textrm{eff}} = (1+4)/2 = 2.5$.

In order to further explore the excitation spectrum, we show the energy shift of nine consecutive Coulomb peaks in a magnetic field applied normal to the plane of the quantum dot in Fig. 3. The vertical energy axis was obtained by converting plunger gate voltage into energy usign the measured lever arm ($\alpha_{\textrm{pg}}=0.075$). In the constant-interaction model, the ground-state energy of an $N$-particle quantum dot can be written as the sum of the single-particle energies $\varepsilon_i(B)$ plus an electrostatic charging energy $NE_C$. The ground-state energy is tuned by the gate voltage $V_g$. The experiment was done in the zero-bias regime; hence we measured the chemical potential of the $N$th Coulomb resonance as explained in \cite{Schnez08}. Experimentally, the single-particle energy $\varepsilon_N(B)$ of the $N$th Coulomb resonance is then determined by $\varepsilon_N(B)=e\alpha_{\textrm{pg}}V_g^{\textrm{res}}(N,B)+NE_C+\textrm{const}$. The constant part and the electrostatic contribution $NE_C$ are subtracted such that consecutive peaks touch each other (alternatingly shown as red triangles and blue circles, respectively). Characteristic lines (see dashed lines in Fig. 3) linear in $B$ with slopes of around $\pm 2.5\,\textrm{meV/T}$ can be seen. This strong $B$-field dependence cannot be explained by the Zeeman effect, which would result in a slope of $g\mu_B = 116\,\mu\textrm{eV/T}$, assuming a $g$-factor of $g = 2$. For higher magnetic fields, the Landau level degeneracy increases and fewer Landau levels are filled. Consequently, the energy spectrum is expected to evolve from single-level fluctuations into a regular pattern. This transition can be seen at around $4\,\textrm{T}$. Recent theoretical calculations are in qualitative agreement with our experimental data \cite{Schnez08, Recher08}.  

\begin{figure}
  \centering
  \includegraphics[width=.45\textwidth]{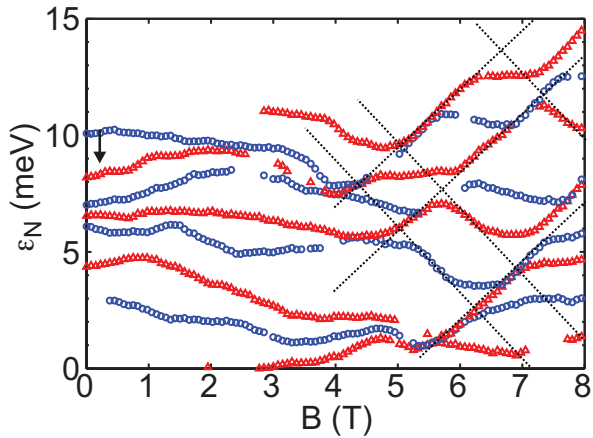}
  \caption{(Color online) Experimental energy spectrum of the quantum dot in a perpendicular magnetic field. The typical magnetic field scale at which a significant change is expected is approximately given by one flux quantum $\Phi_0 = h/e$ per dot area, i.e. $4\Phi_0/\pi d^2 = 270\,\textrm{mT}$ and is indicated by the black arrow. Starting around $B=4\,\textrm{T}$, a regular pattern with characteristic linear slopes evolves (see dashed lines) which shows the transition from single-particle fluctuations to $B$-field dependence.}
\end{figure}

Quantum dots are envisioned as possible building blocks for a future quantum information processor \cite{Loss98}. The preparation and detection of well-defined orbital and spin states are an essential prerequisite for this purpose. On the other hand, graphene might be a highly suitable material for spin manipulations in a condensed-matter environment because of its expected long spin-coherence times \cite{Trauzettel07}. In this paper, we showed that the direct measurement of excited states in graphene quantum dots through transport experiments is possible. This is a first and essential step towards possible experiments with graphene quantum structures.

\emph{Acknowledgement:} We thank K.~S. Novoselov and A.~K. Geim for valuable discussion. Financial support by ETH Z\"urich and the Swiss Science Foundation is gratefully acknowledged.

\end{document}